\documentclass[12pt,a4paper]{article}
\pdfoutput=1
\usepackage{mathrsfs}
\usepackage{amsmath,amsfonts,amssymb,amsbsy,dsfont}
\usepackage{graphicx,color}
\usepackage[nosort]{cite}

\newcommand{\be}{\begin{equation}}
\newcommand{\ee}{\end{equation}}
\newcommand{\bea}{\begin{eqnarray}}
\newcommand{\eea}{\end{eqnarray}}
\newcommand{\ba}{\begin{eqnarray}}
\newcommand{\ea}{\end{eqnarray}}
\newcommand{\nn}{{\nonumber}}
\newcommand{\vev}[1]{{\left< {#1} \right>}}

\newcommand*{\pvint}{
  \mathop{\,\,\vphantom{\intop}\!\!\!
  \mathpalette\pvintop\relax}\nolimits}
\newcommand*{\pvintop}[2]{
  \ooalign{$#1\intop$\cr\hidewidth$#1-$\hidewidth}}

\newcommand{\email}[1]{\vbox{\center\tt#1}\vspace{3mm}}

\begin{document}
\begin{titlepage}

\rightline{\small{\tt }}
\begin{center}
\vskip 1cm
\centerline{{\Large {\bf Ladder exponentiation for generic}}}

~

\centerline{{\Large {\bf large symmetric representation Wilson loops}}}
\vskip 1cm

{Diego H. Correa, Fidel I. Schaposnik Massolo }

{\it Instituto de F\'isica La Plata, CONICET,
Universidad Nacional de La Plata

C.C. 67, 1900 La Plata, Argentina
}
\email{correa@fisica.unlp.edu.ar,\;
fidel.s@fisica.unlp.edu.ar}

\vskip 3cm

\end{center}

\begin{abstract}
A recent proposal was made for a large representation rank limit for which the expectation values of ${\cal N}=4$ super Yang-Mills Wilson loops are given by the exponential of the 1-loop result. We verify the validity of this exponentiation in the strong coupling limit using the holographic D3-brane description for straight Wilson loops following an arbitrary internal space trajectory.
\end{abstract}

\end{titlepage}


\section{Introduction}

A remarkable property of the AdS/CFT correspondence is its weak/strong coupling nature, which allows the use of this duality to study strongly coupled phenomena. However, this property is also the obstacle which prevents from comparing straightforwardly gauge and string theory results. In this regard, the discovery and study of observables that can be exactly described as functions of the coupling constant proves valuable. In some cases, the possibility of obtaining exact results is due to supersymmetry, as happens for instance for the expectation value of circular Wilson loops in ${\cal N}= 4$ super Yang-Mills theory \cite{ESZ,Drukker:2000rr}, which are studied using localization techniques \cite{Pestun}. Exact results for other supersymmetric Wilson loops are also known (see for example \cite{DGRT1,DGRT2,DGRT3,pestun2,Giombi:2009ms,Giombi:2009ds,FT}).

Exact results for certain states which are not supersymmetric have also been obtained. Most of those cases involve a near BPS limit, {\it i.e.} a parametric expansion around some BPS or supersymmetric state. One example is the exact computation of the Bremsstrahlung radiation for a quark in ${\cal N}= 4$ super Yang-Mills \cite{CHMS,Fiol:2012sg,Gromov:2012eu,Fiol:2013iaa,Gromov:2013,Fiol:2015spa}. Another example is the BMN limit for single trace operators \cite{Berenstein:2002jq}. Large R-charge BMN operators are almost BPS and this underlies the successful comparison between gauge and string theory results in that case.

In the article \cite{Correa:2015wma} a limit was proposed to study the expectation value for Wilson loops in large totally symmetric representations of $U(N)$. More precisely, it was observed that when the rank of the representation $k$ is much larger than the rank of the gauge group $N$, ladder diagrams\footnote{Feynman diagrams with no internal vertices.} dominate over the interaction diagrams in the perturbative expansion. Moreover,  the color factors associated with different ladder diagrams are such that the leading contribution to the $\ell$-loop order in this limit turns out to be the $\ell$-th power of the 1-loop ladder diagram and their sum exponentiates the 1-loop result. This led to the conjecture
\be\label{conjecture}
\vev{W_{S_k}} \simeq \exp\langle W_{S_k}^{({\rm 1-loop})} \rangle\,,\qquad{\rm if}\  k\gg N\,.
\ee

If this result is exact, it should also be valid for large values of the 't Hooft coupling, when the expectation value of the Wilson loop should agree with the exponential of the on-shell action of some D3-branes,
\be\label{vevstrongcoupling}
\vev{W_{S_k}} \simeq \exp(-S_{\rm on-shell}^{\rm D3} )\,,
\ee
where one should take $N\ll k\ll N^2$ to ensure no backreaction deforming the ${\rm AdS}_5\times{\rm S}^5$ background takes place.

In \cite{Correa:2015wma} the coincidence between \eqref{conjecture} and \eqref{vevstrongcoupling} in the strong coupling limit for large $k$ was verified for a very particular type of Wilson loop. More specifically, this was shown for a Wilson loop defined over a straight line in spacetime with a cusp in one direction of the internal space. The coincidence can also be observed for circular Wilson loops \cite{HK,FT}.

The main goal of this note is to provide a more sophisticated verification of the coincidence between \eqref{conjecture} and \eqref{vevstrongcoupling} by considering a Wilson line for an {\it arbitrary trajectory} in the internal space.

The most general Wilson loop for a given representation $\mathcal{R}$ is defined by an arbitrary curve ${\cal C}$ in spacetime and an arbitrary trajectory in the internal space given by $\vec n(s)$, which account for the coupling with the gauge potential $A_\mu$ and the scalar fields $\vec\Phi$ respectively,
\be\label{generalvev}
W_{\cal R}({\cal C},\vec n) = \frac{1}{{\rm dim} ({\cal R})} {\rm tr}_{\cal R}\,{\rm P}\!\exp \oint\limits_{\cal C} \left(i A_\mu \dot x^\mu  + |\dot x|  \vec\Phi \cdot\vec n \right)ds\,.
\ee

In this note we consider a Wilson loop in the  $k$-th rank symmetric representation of $U(N)$, \emph{i.e.} we take $\mathcal{R} = S_k$, defined for a straight line in spacetime and an arbitrary trajectory in ${\rm S}^5$ corresponding to an arbitrary unit vector $\vec n(s) \in \mathbb{R}^6$.
\begin{figure}[h!]
\centering
\def\svgwidth{4cm}
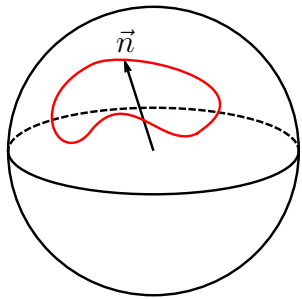
\caption{Internal space trajectory.}
\end{figure}

The expectation value of \eqref{generalvev} is computed perturbatively by expanding the exponential. For the straight line, up to 1-loop order one has
\be
\langle W_{\cal R} \rangle \simeq  1 + \tfrac{{\rm tr}(T^aT^b)}{2\dim{({\cal R})}}\!\iint\!\! dt dt'
\left( n^i(t) n^j(t') \langle \Phi_i^a(t) \Phi_j^b(t')\rangle-
\dot x^\mu(t) \dot x^\nu(t') \langle A^a_\mu(t) A^b_\nu(t')\rangle\right),
\ee
where $T^a$ are the generators in representation ${\cal R}$ and the expectation values are just given by free propagators. Thus,
\be
\langle W_{\cal R}^{({\rm 1-loop})} \rangle = \frac{\lambda}{16\pi^2} \frac{C_2({\cal R})}{N}\!\iint\!
dt dt'\;\frac{1- n^i(t) n^i(t')}{(t-t')^2}\,,
\ee
where $C_2({\cal R})$ is the quadratic Casimir coefficient and $\lambda = g_{\rm YM}^2 N$ is the t'Hooft coupling constant. For totally symmetric representations $C_2(S_k) =  k (N+k-1)$, so
\be\label{W1looplartgek}
\langle W_{S_k}^{({\rm 1-loop})} \rangle \simeq \frac{\lambda}{16\pi^2} \frac{k^2}{N}\!\iint\!
dt dt'\;\frac{1- n^i(t) n^i(t')}{(t-t')^2}\,,\qquad{\rm for }\ k\gg N\,.
\ee

The straight line in $\mathbb{R}^4$ can be equivalently pictured as two antipodal lines along the cylinder  $\mathbb{R}\times {\rm S}^3$, sitting at $\psi=0$ and $\psi = \pi$. This alternative picture is more convenient to compare with D3-brane results obtained for global coordinates. Using the Euclidean time of the cylinder as the curve parameter, the 1-loop expectation value \eqref{W1looplartgek} can be recast as
\be
\langle W_{S_k}^{({\rm 1-loop})} \rangle \simeq \frac{\lambda}{32\pi^2} \frac{k^2}{N}\!\iint\!
d\tau d\tau'\left.\frac{1- n^i_\psi(\tau) n^i_{\psi'}(\tau')}{\cosh(\tau-\tau')-\cos{(\psi-\psi')}}\right|_{\psi,\psi' = 0,\pi}\quad{\rm for }\ k\gg N\,,
\label{W1looplartgekcylinder}
\ee
where $\vec n(t)$ has been split into $\vec n_{0}(\tau)$ and $\vec n_{\pi}(\tau)$. For \eqref{conjecture} and \eqref{vevstrongcoupling} to coincide, \eqref{W1looplartgekcylinder} should agree with minus the on-shell action for the dual D3-brane configurations.

\section{Large $k$ D3-brane}

We take the metric of Euclidean ${\rm AdS}_5 \times {\rm S}^5$ with radius $L$ in the coordinates
\be
\frac{ds^2}{L^2} = \frac{du^2}{1+u^2} + \left(1+u^2\right)d\tau^2 + u^2\left(d\psi^2 + \sin^2\psi \, d\Omega_2^2\right) + d\Omega_5^2\,,
\ee
and write down the action for a D3-brane with asymptotically ${\rm AdS}_2 \times {\rm S}^2$ worldvolume spanning $\zeta^I = (\tau, \psi, \theta, \phi)$, where $\theta$ and $\phi$ parametrize the ${\rm S}^2 \subset {\rm AdS}_5$. We let the angles in the ${\rm S}^5$ be $\varphi_i$ for $i = 1, \cdots, 5$ and make the ansatz $u(\tau, \psi)$ and $\varphi_i(\tau, \psi)$. The DBI part of the action is
\be
S_{\rm DBI} = \frac{N}{2\pi^2} \int d^4\zeta \sqrt{\det\left(\frac{dx^\mu}{d\zeta^I}\frac{dx^\nu}{d\zeta^J}g_{\mu\nu} + \frac{2\pi}{\sqrt\lambda} \, F_{IJ}\right)},
\ee
where we assume the only non-vanishing components of the electromagnetic field are ${F_{\psi\tau} = -F_{\tau\psi} = \partial_\psi A_\tau(\tau,\psi)}$. For the Wess-Zumino term we have
\be
S_{\rm WZ} = -\frac{2N}{\pi} \int d\tau \, d\psi \, \sin^2\psi \, u(\tau, \psi)^4,
\ee
where we already performed the trivial integration over the ${\rm S}^2$. Because the action $S = S_{\rm DBI} + S_{\rm WZ}$ only depends on $\partial_\psi A_\tau(\tau, \psi)$ we can use the equation of motion for $A_\tau(\tau, \psi)$ to introduce the momentum
\be
\frac{1}{N} \frac{\partial \mathcal{L_{\rm DBI+WZ}}}{\partial (\partial_\psi A_\tau)} = i \Pi_A(\tau),
\ee
where the $i$ is put to ensure $\Pi_A(\tau)$ is a real quantity, as the gauge field is imaginary in the Euclidean theory.

We may solve the equation for $A_\tau(\tau, \psi)$ in terms of $\Pi_A(\tau)$, $u(\tau, \psi)$ and $\varphi_i(\tau, \psi)$, and then use this in the remaining equations of motion.
The information about the dual Wilson loop is encoded in electric flux carried by the D3-brane and the boundary conditions of the embedding coordinates \cite{GP}. We shall look for solutions by deforming the well-known 1/2 BPS solution\cite{DF}, so we will take
\be
\label{eq:ansatz_u}
u(\tau, \psi) = \frac{\kappa}{\sin\psi}f(\tau, \psi) \qquad\text{and}\qquad \Pi_A(\tau) = \frac{k}{N}p_A(\tau)\,.
\ee
where $\kappa = k\sqrt{\lambda}/4N$. Note that because we want to relate our solution to a Wilson loop in the symmetric representation of rank $k$, we will immediately take $p_A(\tau) \mapsto 1$. Moreover we are interested in the $k \to \infty$ limit, in which the equations of motion are drastically simplified.
Before we consider the most general case, we will first take a trajectory in some ${\rm S}^1 \subset {\rm S}^5$. For example, if we take $\varphi_1 = \cdots =\varphi_4 =\frac{\pi}{2}$ we have
\be\label{eoms_nobeta}
\Delta f = f |\vec\nabla \varphi_5|^2\,, \qquad\text{and}\qquad f \Delta \varphi_5 + 2 \vec\nabla f \cdot \vec\nabla\varphi_5 = 0\,,
\ee
where $\Delta = \vec\nabla^2$ is the Laplacian operator. Equations \eqref{eoms_nobeta} have to be solved with the following boundary conditions\footnote{In our specific case, there are two boundaries at $\psi = 0$ and $\psi=\pi$, so we would have
\be\nonumber
\varphi_5(\tau, 0) = a_0(\tau)\, \qquad\text{and}\qquad \varphi_5(\tau, \pi) = a_\pi(\tau)\,.
\ee
}
\be
\left.f(\tau, \psi)\right|_{\psi=0,\pi} = 1\,,
\qquad
\left.\varphi_5(\tau, \psi)\right|_{\psi=0,\pi} = a_\psi(\tau)\,,
\ee
for arbitrary functions $a_\psi(\tau)$, which corresponds to the choice
\be
\vec n_\psi(\tau) = (0,0,0,0,\sin a_\psi(\tau),\cos a_\psi(\tau))
\ee
in the dual Wilson loop. To look for solutions of these equations of motion, we perturb around the BPS solution $f=1$ and $\varphi_5 =0$,
\begin{align}
f(\tau, \psi) &= 1 + \epsilon f_1(\tau, \psi) + \epsilon^2 f_2(\tau, \psi) + \cdots\label{expansionf}\\
\varphi_5(\tau, \psi) &= \epsilon\,\phi_1(\tau, \psi) + \epsilon^2 \phi_2(\tau, \psi) + \cdots,
\label{expansiona}
\end{align}
and introduce this expansion in the equations \eqref{eoms_nobeta}. We find, that $f_{2n+1}$ and $\phi_{2n}$ vanish because they have to solve the Laplace equation with vanishing Dirichlet boundary conditions. The non-vanishing $f_{2n}$ and $\phi_{2n-1}$ satisfy
\begin{align}
\Delta f_{2n} &= \sum_{j=0}^{n-1} f_{2j} \sum_{i=1}^{n-j} \vec\nabla\phi_{2i-1} \cdot \vec\nabla\phi_{2(n-j) - (2i-1)}\label{eq:nabla_f}\,,\\
\Delta \phi_{2n-1} &= - \sum_{i=1}^n 2\vec\nabla f_{2i}\cdot\vec\nabla\phi_{2n-2i+1} + f_{2i} \Delta \phi_{2n-2i+1}\label{eq:nabla_a}\,,
\end{align}
where we defined $f_0(\tau, \psi) = 1$ for convenience. In order to find solutions to the system \eqref{eq:nabla_f}-\eqref{eq:nabla_a}, it is useful
to define functions $g_n$ as
\be
g_n =
\begin{cases}
f_n & \text{if $n$ is even}\,,
\\
i \phi_n &  \text{if $n$ is odd}\,,
\end{cases}
\label{defgn}
\ee
which allows us to write the equation for the $m$-th coefficient $g_m$ in terms of some ordered, restricted partitions $P^{1{\rm e}}_m$ of the number $m$,
\be
\Delta \left(\sum_{\mathcal{Y} \in P^{1{\rm e}}_m} \prod_{y_i \in \mathcal{Y}} g_{y_i}\right) = 0\,,
\label{laplace partition_ordered}
\ee
where the elements of $P^{1{\rm e}}_m$ contain at most one even element. By excluding the trivial partition, a solution for the coefficient $g_m$ is given in terms of the previous ones,
\be
g_m = -\sum_{\mathcal{Y} \in {P^{1{\rm e}}_m}'}  \prod_{y_i \in \mathcal{Y}} g_{y_i}\,,
\ee
where the sum is now over the non-trivial ordered partitions ${P^{1{\rm e}}_m }'$, with at most one even element. The restricted partitions and their corresponding coefficients can be conveniently packed in the following generating function
\begin{align}
Z(\epsilon)
&= \sum_{k=0}^\infty {g}_{2k} \epsilon^{2k} \prod_{n=1}^\infty
\left(\sum_{l=0}^\infty \frac{1}{l!} \left(g_{2n-1} \epsilon^{2n-1}\right)^l\right)
\nn
\\
& =
\sum_{k=0}^\infty {f}_{2k} \epsilon^{2k}
\exp\left( i\sum_{n=1}^\infty \phi_{2n-1}\epsilon^{2n-1}\right)
= f e^{i\varphi_5}
\,.
\end{align}
Due to \eqref{laplace partition_ordered}, the Laplacian of the generating function is vanishing. This means that expanding around the BPS solution was unnecessary, since with a simple change of variables the system \eqref{eoms_nobeta} could have been linearized and decoupled. Indeed, if we define the complex function $Z(\tau, \psi) = f(\tau, \psi) \, e^{i\varphi_5(\tau, \psi)}$ we see that
\be
\Delta Z = e^{i\varphi_5} \left(\Delta f - f |\vec\nabla\varphi_5|^2\right) + i\,e^{i\varphi_5}\left(f\Delta\varphi_5 + 2\vec\nabla f \cdot \vec\nabla\varphi_5\right)\,.
\ee
Thus, $\Delta Z = 0$ is equivalent to the original system \eqref{eoms_nobeta}. To solve the Laplace equation with Dirichlet boundary conditions, \emph{i.e.}
\be
\label{dirichlet problem}
\Delta Z = 0 \qquad\text{with}\qquad \left.Z\right|_{\psi=0,\pi} = e^{i\,a_\psi(\tau)}\,,
\ee
one can use the Green function method. In our case, where the two dimensional domain is a strip with boundaries at $\psi=0$ and $\psi=\pi$, we shall use the Green function
\be
G(\tau, \psi; \tau', \psi') = \frac{1}{4\pi} \log\frac{\cosh(\tau-\tau')-\cos(\psi+\psi')}{\cosh(\tau-\tau') - \cos(\psi-\psi')}\,.
\ee
Thus, the solution to this Dirichlet problem is
\be
Z(\tau, \psi) = \int d\tau' \left.\vphantom\sum\partial_\perp' G(\tau, \psi; \tau', \psi') e^{i\,a_{\psi'}(\tau')} \right|_{\psi'=0,\pi},
\ee
where $\partial_\perp$ denotes the normal (to the boundary) outward derivative, so that $\partial_\perp' = -\partial_{\psi'}$ at $\psi'=0$ and $\partial_\perp' = +\partial_{\psi'}$ at $\psi'=\pi$.

Armed with the intuition provided by the ${\rm S}^1$ case, we may now proceed to solve the case corresponding to a general trajectory in the ${\rm S}^5$. Introducing $\chi^a = (f, \varphi_1, \varphi_2, \varphi_3, \varphi_4, \varphi_5),$ the equations of motion in the large $k$ limit can be compactly written as
\be
\Delta \chi^a = \Gamma^a_{\hphantom{a}bc} \vec\nabla\chi^b\cdot\vec\nabla\chi^c\,,
\label{eomsinspherical}
\ee
where $\Gamma^a_{\hphantom{a}bc}$ are the Christoffel symbols corresponding to $\mathbb{R}^6$ in spherical coordinates with $f$ as the radius. Introducing cartesian coordinates $X^i$
\begin{align}
X^1(\tau, \psi) &= f(\tau, \psi) \sin\varphi_1(\tau, \psi) \sin\varphi_2(\tau, \psi) \sin\varphi_3(\tau,\psi)\sin\varphi_4(\tau, \psi)\sin\varphi_5(\tau,\psi)\,,\nonumber\\
X^2(\tau, \psi) &= f(\tau, \psi) \sin\varphi_1(\tau, \psi) \sin\varphi_2(\tau, \psi) \sin\varphi_3(\tau,\psi) \sin\varphi_4(\tau, \psi)\cos\varphi_5(\tau,\psi)\,,\nonumber\\
X^3(\tau, \psi) &= f(\tau, \psi) \sin\varphi_1(\tau, \psi) \sin\varphi_2(\tau, \psi) \sin\varphi_3(\tau,\psi) \cos\varphi_4(\tau,\psi)\,,\nonumber\\
X^4(\tau, \psi) &= f(\tau, \psi) \sin\varphi_1(\tau, \psi) \sin\varphi_2(\tau,\psi) \cos\varphi_3(\tau,\psi)\,,\\
X^5(\tau, \psi) &= f(\tau, \psi) \sin\varphi_1(\tau, \psi) \cos\varphi_2(\tau,\psi)\,,\nonumber\\
X^6(\tau, \psi) &= f(\tau, \psi) \cos\varphi_1(\tau, \psi)\,,\nonumber
\end{align}
the equations \eqref{eomsinspherical} become
\be\label{eomsincartesian}
\Delta X^i = 0\,,
\ee
which have to be solved with boundary conditions
\be
\left.\vphantom\sum X^i(\tau,\psi)\right|_{\psi=0,\pi} = n_\psi^i(\tau)\,.
\ee
Thus, in this large $k$ limit the solution for an arbitrary internal space trajectory is given by
\be\label{solucion}
X^i(\tau, \psi) = \int d\tau' \left.\vphantom\sum\partial_\perp' G(\tau, \psi; \tau', \psi') n_{\psi'}^i(\tau') \right|_{\psi'=0,\pi}.
\ee

\section*{On-shell action}

We would like to evaluate the action on-shell. In addition to the bulk action terms $S_{\rm DBI}$ and $S_{\rm WZ}$ there will be boundary terms of the form \cite{DF,DGO}
\be
S_{\rm bdry} = -\int \left.\vphantom\sum d\tau\left(u \frac{\partial \mathcal{L_{\rm DBI+WZ}}}{\partial (\partial_\psi u)} + A_\tau \frac{\partial \mathcal{L_{\rm DBI+WZ}}}{\partial (\partial_\psi A_\tau)}\right)\right|_{\psi=0}^{\psi=\pi}\,.
\ee
The boundary term for $A_\tau$ can be rewritten as an integral over the bulk
\be
\int d\tau \left.\vphantom\sum A_\tau(\tau, \psi) \frac{\partial \mathcal{L_{\rm DBI+WZ}}}{\partial (\partial_\psi A_\tau)}\right|_{\psi=0}^{\psi=\pi} = N i \int d\tau \, d\psi \, \partial_\psi A_\tau(\tau, \psi) \, \Pi_A(\tau)\,.
\ee
If we perform a similar trick with the boundary term for $u$, we can write the full on-shell action as an integration over the bulk. Once all the contributions are brought together, the on-shell action in the $k\to\infty$ limit reads
\begin{align}
S_{\rm on-shell} &= \frac{k^2 \lambda}{16\pi N}\int d\tau d\psi \left(\frac{1 - f^2}{\sin^2\psi} - |\vec\nabla f|^2 - 2 f \Delta f
+ f^2 \sum_{i=1}^5 |\vec\nabla \varphi_i|^2 \prod_{j=1}^{i-1}\sin^2\varphi_j\right.\nonumber\\
&\qquad\qquad\qquad\qquad\quad\left.\vphantom{\frac{1}{2}}+ 2\left(\partial_\tau f\right)^2 + 2f \left(\cot\psi \partial_\psi f + \partial^2_\tau f\right)\right).
\end{align}
Using the equation of motion for $f$, which states that
\be
\Delta f = f \sum_{i=1}^5 |\vec\nabla \varphi_i|^2 \prod_{j=1}^{i-1}\sin^2\varphi_j\,,
\ee
we can rewrite this as
\be
S_{\rm on-shell} = \frac{k^2 \lambda}{16\pi N}\int d\tau d\psi \left(\tfrac{1}{2}\Delta f^2 + \partial_\psi\left[\cot\psi (f^2-1) - \partial_\psi f^2\right]\right)\,.
\ee
The second term in the above expression gives upon integration a boundary term which vanishes when we take $f(\tau, \psi) \to 1$ at the boundary. If we further use that ${X^iX^i} = f^2$, we then see that the on-shell action is obtained by evaluating the integral
\be
S_{\rm on-shell} = \frac{k^2 \lambda}{32\pi N}\int d\tau \, d\psi \, \Delta (X^iX^i) = \frac{k^2 \lambda}{16\pi N}\int d\tau \left.\vphantom\sum X^i \partial_\perp X^i\right|_{\psi=0,\pi}\,.
\ee
Using our representation \eqref{solucion}, the on-shell action takes the form of a double integral over the boundary,
\begin{align}
S_{\rm on-shell} &= \frac{k^2 \lambda}{16 \pi N}\iint d\tau d\tau' \left.\vphantom\sum \partial_\perp \partial_\perp' G(\tau, \psi; \tau', \psi')  
n_\psi^i(\tau) n_{\psi'}^i(\tau') \right|_{\psi,\psi'=0,\pi}
\\
&= -\frac{k^2 \lambda}{64\pi^2 N}\iint d\tau d\tau' \left.\frac{\left[n_{\psi}^i(\tau) - n_{\psi'}^i(\tau')\right]^2}{\cosh(\tau-\tau') - \cos(\psi-\psi')}\right|_{\psi,\psi'=0,\pi}
\end{align}
where we used the Green function property presented in the Appendix. Finally, since $n^i$ is a unit vector, the on-shell action takes the form
\begin{align}
S_{\rm on-shell}&= -\frac{k^2 \lambda}{32\pi^2 N}\iint d\tau d\tau' \left.\frac{1-n_\psi^i(\tau)n_{\psi '}^i(\tau')}{\cosh(\tau-\tau') - \cos(\psi-\psi')}\right|_{\psi,\psi'=0,\pi}\,,
\end{align}
which is exactly minus the 1-loop contribution to the Wilson loop expectation value in the large rank limit given in \eqref{W1looplartgekcylinder}.

The agreement between the on-shell action and the 1-loop expectation value in the large $k$ limit is indicating that the ladder exponentiation proposed in \cite{Correa:2015wma} is correct. We would like to stress that although the spacetime trajectory of the Wilson loop considered is a straight line, the internal space trajectory is kept completely arbitrary. As a result, the verification of the ladder exponentiation is more general that the one found in \cite{Correa:2015wma}, where the Wilson loop had an internal cusp which corresponds to the case of a step function in an ${\rm S}^1\subset {\rm S}^5$. This strengthens the idea that the ladder exponentiation in the large rank limit takes place for arbitrary Wilson loops. Note that this exact result to leading order in the proposed large rank limit is not related to supersymmetry. The large rank limit does not seem to be in any obvious way a near BPS limit.

The ladder exponentiation resembles the Abelian eikonal exponentiation of QED, according to which the vacuum expectation value of a Wilson loop is exactly the exponential of the 1-loop result, given by the photon propagator \cite{Yennie:1961ad}. For non-Abelian theories, the vacuum expectation value of a Wilson loop is the exponential of a more complicated expression to which not only the 1-loop diagram contributes. Given the fact that the Abelian part exponentiates it is possible to write this expression in terms of a smaller set of diagrams known as webs \cite{Sterman:1981jc,Gatheral:1983cz,Frenkel:1984pz}. However, in our limit the color factors of all these terms are such that their contribution to $\langle W \rangle$ is subleading with respect to the Abelian part involving powers of the quadratic Casimir.


\subsection*{Acknowledgements}
We would like to thank Mariel Santangelo and Bartomeu Fiol for helpful comments. The authors are supported by CONICET and grants PICT 2012-0417 and PIP 0681.

\section*{Appendix}\label{appendix}

To evaluate the on-shell action we will need to perform integrals of the form
\be
\mathcal{I}[u,v] = \iint d\tau d\tau' u_\psi(\tau) v_{\psi'}(\tau') \left.\vphantom\sum \partial_\perp \partial_\perp' G(\tau, \psi; \tau', \psi')\right|_{\psi,\psi'=0,\pi}.
\ee
These split into four cases depending on the choice of $\psi, \psi' = 0, \pi$, so it is convenient to perform a change of variables to join them into a single integration. We will take
\be
x = e^{\tau}\cos\psi \qquad\text{and}\qquad y = e^{\tau}\sin\psi,
\ee
so that we map the strip $(\tau,\psi)\in\mathbb{R}\times[0,\pi]$ into the half-plane $y>0$. We then have
\be
\mathcal{I}[u,v] =  \iint dx\,dx' u(x) v(x') \left.\vphantom\sum \partial_y \partial_{y'} \tilde{G}(x,y; x',y')\right|_{y,y'=0}\,,
\ee
In these new variables, the Green function has the simpler form
\be
\tilde{G}(x,y; x',y') = \frac{1}{4\pi} \log\frac{(x-x')^2 + (y+y')^2}{(x-x')^2 + (y-y')^2}\,,
\ee
so that
\be
\mathcal{I}[u,v] =  \frac{1}{\pi}\lim_{y\to0} \iint dx\,dx' u(x) v(x') \frac{(x-x')^2 - y^2}{\left( (y)^2 + (x-x')^2 \right)^2}\,.
\ee
At this point it would be incorrect to take the $y\to0$ limit before integrating because it would lead to an integrand with a double pole, which would be unphysical. Thus, before taking the limit we shall write the integrand as a principal value integral. It is easy to see that
\be
\mathcal{I}[u,v] =  \frac{1}{\pi^2}\lim_{y\to0} \iint dx\,dx' u(x) v(x')
\pvint\limits_{z\neq x} \!\! \frac{dz}{(z-x)^2}\left(\frac{y}{y^2+(z-x')^2}-\frac{y}{y^2+(x-x')^2}\right)\,.\nn
\ee
We further rephrase this by introducing another integral with a delta function and replicating the expression
\begin{align}
\mathcal{I}[u,v]  &= \lim_{y\to0} \iint\!\! dx\,dx' u(x) v(x')\!
\int \!\!dz'\!\!\pvint\limits_{z\neq z'} \!\! \frac{dz\delta(z'-x)}{2\pi^2(z-z')^2}\left(\frac{y}{y^2+(z-x')^2}-\frac{y}{y^2+(z'-x')^2}\right)\nn
\\
&+\lim_{y\to0} \iint\!\! dx\,dx' u(x) v(x')\!
  \int \!\!dz\!\!\pvint\limits_{z'\neq z} \!\! \frac{dz'\delta(z-x)}{2\pi^2(z-z')^2}\left(\frac{y}{y^2+(z'-x')^2}-\frac{y}{y^2+(z-x')^2}\right)\nn
\end{align}
If we now take the limit inside the integrals we obtain four products of delta functions whose combination results in a convergent integral, even
without taking principal values
\begin{align}
\mathcal{I}[u,v]  &= \!\iint\!\! dx\,dx' u(x) v(x')\!\iint \!\!dz dz'\frac{(\delta(z-x)-\delta(z'-x))(\delta(z'-x')-\delta(z-x'))}{2\pi(z-z')^2}\nn
\\
&=  \iint \!\!dz dz'\frac{(u(z)-u(z'))(v(z')-v(z))}{2\pi(z-z')^2}\,.
\end{align}

We can express this result in terms of the original coordinates. For $z>0$ we have $z=e^\tau$ that corresponds to the line at $\psi = 0$, while  for $z<0$ we have $z=-e^\tau$. Thus,
\be
\mathcal{I}[u,v] =  - \frac{1}{4\pi}\iint \!\!d\tau d\tau'\left.\frac{(u_{\psi}(\tau)-u_{\psi'}(\tau'))(v_{\psi}(\tau)-v_{\psi'}(\tau'))}{\cosh(\tau-\tau')-\cos(\psi-\psi')}\right|_{\psi,\psi'=0,\pi}.
\ee


\begin{thebibliography}{99}
\addtolength{\parskip}{-.5ex}

\bibitem{ESZ}
  J.~K.~Erickson, G.~W.~Semenoff and K.~Zarembo,
  ``Wilson loops in N=4 supersymmetric Yang-Mills theory,''
  Nucl.\ Phys.\ B {\bf 582}, 155 (2000)
  [hep-th/0003055].

\bibitem{Drukker:2000rr}
  N.~Drukker and D.~J.~Gross,
``An Exact prediction of N=4 SUSYM theory for string theory,''
  J.\ Math.\ Phys.\  {\bf 42} (2001) 2896
  [hep-th/0010274].

\bibitem{Pestun}
  V.~Pestun,
  ``Localization of gauge theory on a four-sphere and supersymmetric Wilson loops,''
  Commun.\ Math.\ Phys.\  {\bf 313}, 71 (2012)
  [arXiv:0712.2824 [hep-th]].

\bibitem{DGRT1}
  N.~Drukker, S.~Giombi, R.~Ricci and D.~Trancanelli,
  ``More supersymmetric Wilson loops,''
  Phys.\ Rev.\ D {\bf 76}, 107703 (2007)
  [arXiv:0704.2237 [hep-th]].

\bibitem{DGRT2}
  N.~Drukker, S.~Giombi, R.~Ricci and D.~Trancanelli,
  ``Wilson loops: From four-dimensional SYM to two-dimensional YM,''
  Phys.\ Rev.\ D {\bf 77}, 047901 (2008)
  [arXiv:0707.2699 [hep-th]].


\bibitem{DGRT3}
  N.~Drukker, S.~Giombi, R.~Ricci and D.~Trancanelli,
  ``Supersymmetric Wilson loops on S**3,''
  JHEP {\bf 0805}, 017 (2008)
  [arXiv:0711.3226 [hep-th]].

\bibitem{pestun2}
  V.~Pestun,
  ``Localization of the four-dimensional N=4 SYM to a two-sphere and 1/8 BPS Wilson loops,''
  arXiv:0906.0638 [hep-th].

\bibitem{Giombi:2009ms}
  S.~Giombi, V.~Pestun and R.~Ricci,
  ``Notes on supersymmetric Wilson loops on a two-sphere,''
  JHEP {\bf 1007}, 088 (2010)
  [arXiv:0905.0665 [hep-th]].


\bibitem{Giombi:2009ds}
  S.~Giombi and V.~Pestun,
  ``Correlators of local operators and 1/8 BPS Wilson loops on S**2 from 2d YM and matrix models,''
  JHEP {\bf 1010}, 033 (2010)
  [arXiv:0906.1572 [hep-th]].

\bibitem{FT}
  B.~Fiol and G.~Torrents,
  ``Exact results for Wilson loops in arbitrary representations,''
  JHEP {\bf 1401}, 020 (2014)
  [arXiv:1311.2058 [hep-th]].

\bibitem{CHMS}
  D.~Correa, J.~Henn, J.~Maldacena and A.~Sever,
  ``An exact formula for the radiation of a moving quark in N=4 super Yang Mills,''
  JHEP {\bf 1206}, 048 (2012)
  [arXiv:1202.4455 [hep-th]].

\bibitem{Fiol:2012sg}
B.~Fiol, B.~Garolera and A.~Lewkowycz,
``Exact results for static and radiative fields of a quark in N=4 super Yang-Mills,''
JHEP {\bf 1205} (2012) 093
[arXiv:1202.5292 [hep-th]].

\bibitem{Gromov:2012eu}
  N.~Gromov and A.~Sever,
  ``Analytic Solution of Bremsstrahlung TBA,''
  JHEP {\bf 1211}, 075 (2012)
  [arXiv:1207.5489 [hep-th]];

\bibitem{Fiol:2013iaa}
  B.~Fiol, B.~Garolera and G.~Torrents,
  ``Exact momentum fluctuations of an accelerated quark in N=4 super Yang-Mills,''
  JHEP {\bf 1306}, 011 (2013)
  [arXiv:1302.6991 [hep-th]].

\bibitem{Gromov:2013}
   N.~Gromov, F.~Levkovich-Maslyuk and G.~Sizov,
  ``Analytic Solution of Bremsstrahlung TBA II: Turning on the Sphere Angle,''
  JHEP {\bf 1310}, 036 (2013)
  [arXiv:1305.1944 [hep-th]].

\bibitem{Fiol:2015spa}
  B.~Fiol, E.~Gerchkovitz and Z.~Komargodski,
  ``The Exact Bremsstrahlung Function in N=2 Superconformal Field Theories,''
  arXiv:1510.01332 [hep-th].

\bibitem{Berenstein:2002jq}
  D.~E.~Berenstein, J.~M.~Maldacena and H.~S.~Nastase,
  ``Strings in flat space and pp waves from N=4 superYang-Mills,''
  JHEP {\bf 0204} (2002) 013
  [hep-th/0202021].

  \bibitem{Correa:2015wma}
  D.~H.~Correa, F.~I.~S.~Massolo and D.~Trancanelli,
  ``Cusped Wilson lines in symmetric representations,''
  JHEP {\bf 1508} (2015) 091
  [arXiv:1506.01680 [hep-th]].

 \bibitem{HK}
  S.~A.~Hartnoll and S.~P.~Kumar,
  ``Higher rank Wilson loops from a matrix model,''
  JHEP {\bf 0608}, 026 (2006)
  [hep-th/0605027].

\bibitem{GP}
  J.~Gomis and F.~Passerini,
  ``Holographic Wilson Loops,''
  JHEP {\bf 0608}, 074 (2006)
  [hep-th/0604007];
  ``Wilson Loops as D3-Branes,''
  JHEP {\bf 0701}, 097 (2007)
  [hep-th/0612022].

\bibitem{DF}
  N.~Drukker and B.~Fiol,
  ``All-genus calculation of Wilson loops using D-branes,''
  JHEP {\bf 0502}, 010 (2005)
  [hep-th/0501109].

\bibitem{DGO}
  N.~Drukker, D.~J.~Gross and H.~Ooguri,
  ``Wilson loops and minimal surfaces,''
  Phys.\ Rev.\ D {\bf 60}, 125006 (1999)
  [hep-th/9904191].

\bibitem{Yennie:1961ad}
  D.~R.~Yennie, S.~C.~Frautschi and H.~Suura,
  ``The infrared divergence phenomena and high-energy processes,''
  Annals Phys.\  {\bf 13} (1961) 379.

\bibitem{Sterman:1981jc}
 G.~F.~Sterman,
  ``Infrared divergences in perturbative QCD,''
  AIP Conf.\ Proc.\  {\bf 74} (1981) 22.

\bibitem{Gatheral:1983cz}
  J.~G.~M.~Gatheral,
  ``Exponentiation of Eikonal Cross-sections in Nonabelian Gauge Theories,''
  Phys.\ Lett.\ B {\bf 133} (1983) 90.

\bibitem{Frenkel:1984pz}
  J.~Frenkel and J.~C.~Taylor,
  ``Nonabelian Eikonal Exponentiation,''
  Nucl.\ Phys.\ B {\bf 246} (1984) 231.

\end{thebibliography}
\end{document}